\documentclass[10pt,a4]{article} 
\usepackage{makeidx}

\newif\ifpdf
\ifx\pdfoutput\undefined
   \pdffalse
\else
   \pdfoutput=1
   \pdftrue
\fi

\ifpdf
  \usepackage[
  pdftex,
  hyperindex,
  pagebackref,
  colorlinks,
  linkcolor=blue,
  menucolor=blue,
  pagecolor=blue,
  urlcolor=blue,
  citecolor=blue,
  pdftitle={The GNSS-R Eddy Experiment I: altimetry from low altitude aircraft},
  pdfauthor={Marco Caparrini},
  pdfcreator={LaTeX}
  ]{hyperref}

  \usepackage{thumbpdf}
  \usepackage{url}
  \usepackage[pdftex]{graphicx}
  \DeclareGraphicsExtensions{.pdf,.jpg,.mps,.png}

\else

  \usepackage[dvips]{graphicx}
  \DeclareGraphicsExtensions{.eps,.jpg,.mps,.png,.ps}

\fi

\usepackage{fancyhdr}  
\usepackage{xfig}
\usepackage{multicol}

\graphicspath{{figures/}}

\usepackage[english]{babel} 
\usepackage[]{times}

\catcode`\@=11

\catcode`\@=11
\def\section{\@startsection{section}{1}{\z@}{3.5ex plus 1ex minus
   .2ex}{2.3ex plus .2ex}{\large\bf}}

\def\eqnarray{\let\@currentlabel=\theequation\refstepcounter{equation}
    \global\@eqnswtrue
    \global\@eqcnt\z@\tabskip\@centering\let\\=\@eqncr
    $$\halign to \displaywidth\bgroup\@eqnsel\hskip\@centering
      $\displaystyle\tabskip\z@{##}$&\global\@eqcnt\@ne
       \hfil${{}##{}}$\hfil
      &\global\@eqcnt\tw@ $\displaystyle\tabskip\z@{##}$\hfil
       \tabskip\@centering&\llap{##}\tabskip\z@\cr}
\def\lefteqn#1{\hbox to 4\arraycolsep{$\displaystyle #1$\hss}}

\def\thesection{\arabic{section}.}

\def\appendix{\setcounter{section}{0}
        \def\thesection{\Alph{section}.}
        \def\theequation{\Alph{section}.\arabic{equation}}}

\long\def\@makefntext#1{\parindent 0cm\noindent
\hbox to 1em{\hss$^{\@thefnmark}$}#1}

\newcommand\unmarkfootnote[1]{%
  \begingroup 
    \let\@makefntext\noindent
    \footnotetext{#1}%
  \endgroup}

\newcommand{\captionfonts}{\footnotesize}

\makeatletter  
\long\def\@makecaption#1#2{%
  \vskip\abovecaptionskip
  \sbox\@tempboxa{{\captionfonts #1: #2}}%
  \ifdim \wd\@tempboxa >\hsize
    {\captionfonts #1: #2\par}
  \else
    \hbox to\hsize{\hfil\box\@tempboxa\hfil}%
  \fi
  \vskip\belowcaptionskip}
\makeatother   

\begin{document}

\thispagestyle{empty}

\noindent
{\Large \bf The GNSS-R Eddy Experiment I: Altimetry from Low Altitude Aircraft} 

\vspace*{0.3cm}
\noindent 
G. Ruffini, F. Soulat, M. Caparrini, O. Germain\\
 {\it Starlab, C. de l'Observatori Fabra s/n, 08035 Barcelona, Spain, http://starlab.es} 
\medskip\\
\noindent
M. Mart\'{i}n-Neira\\
  {\it ESA/ESTEC, Keplerlaan 1, 2200 Noordwijk, The Netherlands,  http://esa.int}

\vspace{0.8cm}
\section*{Abstract}
We report results from the Eddy Experiment, where a synchronous GPS receiver pair was flown on an aircraft to collect sampled L1 signals and their reflections from the sea surface to investigate the altimetric accuracy  of GNSS-R. During the experiment, surface wind speed (U10) was of the order of 10~m/s, and significant wave heights of up to 2~m, as discussed further in a companion paper. After software tracking of the two signals through  despreading of the GPS codes, a parametric waveform model containing the description of the sea surface conditions has been used to fit the waveforms (retracking) and estimate the temporal lapse between the direct GPS signals and their reflections. The estimated lapses have then been used to estimate the sea surface height (SSH) along the aircraft track  using a differential geometric model.  As expected, the precision of GNSS-R ranges was of 3 m after 1 second integration. More importantly, the  accuracy of the GNSS-R altimetric solution with respect to Jason-1 SSH  and {\em in situ}  GPS buoy measurements was of 10 cm, which was the  target with the used experimental setup. This new result confirms the potential of GNSS-R for mesoscale altimetric monitoring of the ocean, and provides an important  milestone on the road to a space mission.
\medskip\\
{\bf Keywords:} GNSS-R, GPS-R, altimetry, meosocale, PARIS, waveform, retracking, bistatic.
\vspace{0.8cm}

\begin{multicols}{2}

\section{Introduction}
Several Global Navigation Satellite System (GNSS) constellations and augmentation systems are presently operational or under development, such as the pioneering US Global Positioning System (GPS),  the Russian Global Navigation Satellite System (GLONASS) and the European EGNOS. In the next few years, the European Satellite Navigation System (Galileo) will be deployed, and GPS will be upgraded with more frequencies and civilian codes. By the time Galileo becomes operational in 2008, more than 50 GNSS satellites will be emitting very precise L-band spread spectrum signals, and will remain in operation for at least a few decades. Although originally meant for (military) localization, these signals will no doubt be used within GCOS/GOOS\footnote{Global Climate Observing System/Global Ocean Observing System.} in many ways (e.g., atmospheric sounding). We focus here on the budding field known as GNSS Reflections, which aims at providing tools for remote sensing of the ocean surface (especially sea surface height and roughness) and the atmosphere over the oceans. 
 
This paper reports a new development of the GNSS-R altimetric concept (PARIS). The PARIS concept (Passive Reflectometry Interferometric System \cite{martin-neira1993})  addresses the exploitation of reflected Global Navigation Satellite Systems signals for altimetry over the oceans. 
Ocean altimetry, the measurement of Sea Surface Height (SSH), is indeed one of the main applications of the GNSS-R passive radar concept. GNSS-R can give automatically integrated measurements in the GNSS reference system. In addition,  this technique can provide the unprecedented spatio-temporal samplings needed for mesoscale monitoring of ocean circulation. It is at mesoscale where phenomena such as eddies play a fundamental role in the transport of energy and momentum, yet current systems are unable to probe them.\\

Many GNSS-R altimetry and scatterometry experiments have been carried out to date, and the list continues to grow thanks to dedicated efforts in Europe and the US. GNSS-R experimental data has now been gathered from  Earth fixed receivers (\cite{caparrini1998,martin-neira2001,treuhaft2001,caparrini2003} among others),  aircraft (\cite{komjathy1998,garrison1998,cardellach2001a,lowe2002,germain2003} among others), stratospheric balloons (\cite{garrison2000,cardellach2003} among others), and from space platforms (\cite{lowe2002b} among others). This experimental work is converging to a unified understanding of the GNSS-R error budget, but so far these experiments have focused on waveform modeling and short term ranging precision. None to date have attempted to retrieve a mesoscale altimetric profile as provided by monostatic radar altimeters such as Jason-1.

In the four main sections of this paper we report PARIS altimetric processing results using data from the 09-27-2002 Eddy Experiment, carried out in the frame of the European Space Agency ``PARIS Gamma'' contract. The first section addresses the issue of {\em tracking} the direct and reflected GPS signals, which consist in appropriately placing the delay and Doppler gating windows and in despreading the GPS signals by means of correlation with clean replicas. Tracking produces  incoherently averaged {\em waveforms} (typically with a cadence of 1~second). The extraction of the information needed for the altimetric algorithm from the waveforms is described in the second section. This is the {\em retracking} step, and it yields to the so-called  {\em measured temporal lapse} (or lapse, for short) between the direct and reflected signal. In the third section, the altimetric algorithm  (producing the Sea Surface Height estimates) is described and, finally, results are presented in the fourth section. 

\section{Data collection and pre-processing}
\subsection{Data set}
\begin{figure*}[t!]
\centering
        \includegraphics[width=10cm]{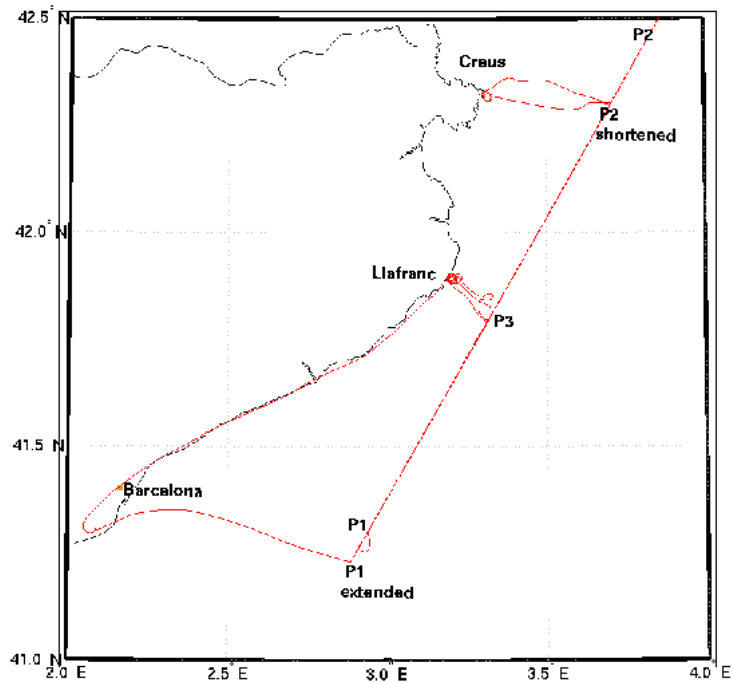} 
    \caption{Flight trajectory.}
  \label{fig:flight}
\end{figure*}

The GNSS-R data set was gathered during an  airborne campaign carried out by Starlab in September~2002. The GPS/INS (Inertial Navigation System) equipped aircraft overflew the Mediterranean Sea, off the coast of Catalonia (Spain), northwards from  the city of Barcelona for about 150~km (Figure~\ref{fig:flight}). This area was chosen because it is crossed by a ground track of the Jason-1 altimeter (track number~187). The aircraft  overflew  this track during the Jason-1 overpass, for precise comparison. In addition, a  GPS buoy measurement of the SSH on a point along the same track was obtained, basically to validate the Jason-1 measurement\footnote{The buoy campaign was carried out by the Institut d'Estudis Espacials de Catalunya (IEEC).}. 

During the experiment, the aircraft overflew twice  the Jason-1 ground track, both in the South-North direction and back, gathering about 2 hours of raw GPS-L1~IF data sampled at 20.456~MHz (see \cite{soulat2003d} for the experimental setup). In this paper, we focus on the data processing of the ascending track, i.e., from P1 to P2  in Figure~\ref{fig:flight}, with the Jason-1 overpass happening roughly in the middle of the track.

\subsection{Tracking GNSS-R signals}
Altimetry with GNSS-R signals is based on the measurement of the temporal {\em lapse} between the time of arrival of the direct GNSS signal and the time of arrival of the same signal after its reflection from the target surface. Successful tracking of both signals is  the first step for altimetric processing.

Under general sea conditions, GPS signals reflected from a rough sea surface cannot be tracked by a standard receiver, because of the signal corruption due to the reflection process itself (in GPS terminology, the signal is affected by severe multipath from the sea clutter). For this reason, a dedicated software receiver has been developed\footnote{STARLIGHT, also described in \cite{caparrini2003}.}. The high level block diagram of this receiver is shown in Figure~\ref{fig:track_block}. The processor is composed of two sub-units, one for each signal. The unit which processes the direct signal---~the master unit~---uses standard algorithms to track the correlation peak of the signal, both in time and frequency. The unit which processes the reflected signal---~the slave unit~---performs correlations blindly, with  time delay and frequency settings which depend on those of the  master unit.

\ifpdf
\begin{figure*}[t!]
\centering
        \includegraphics[width=10cm]{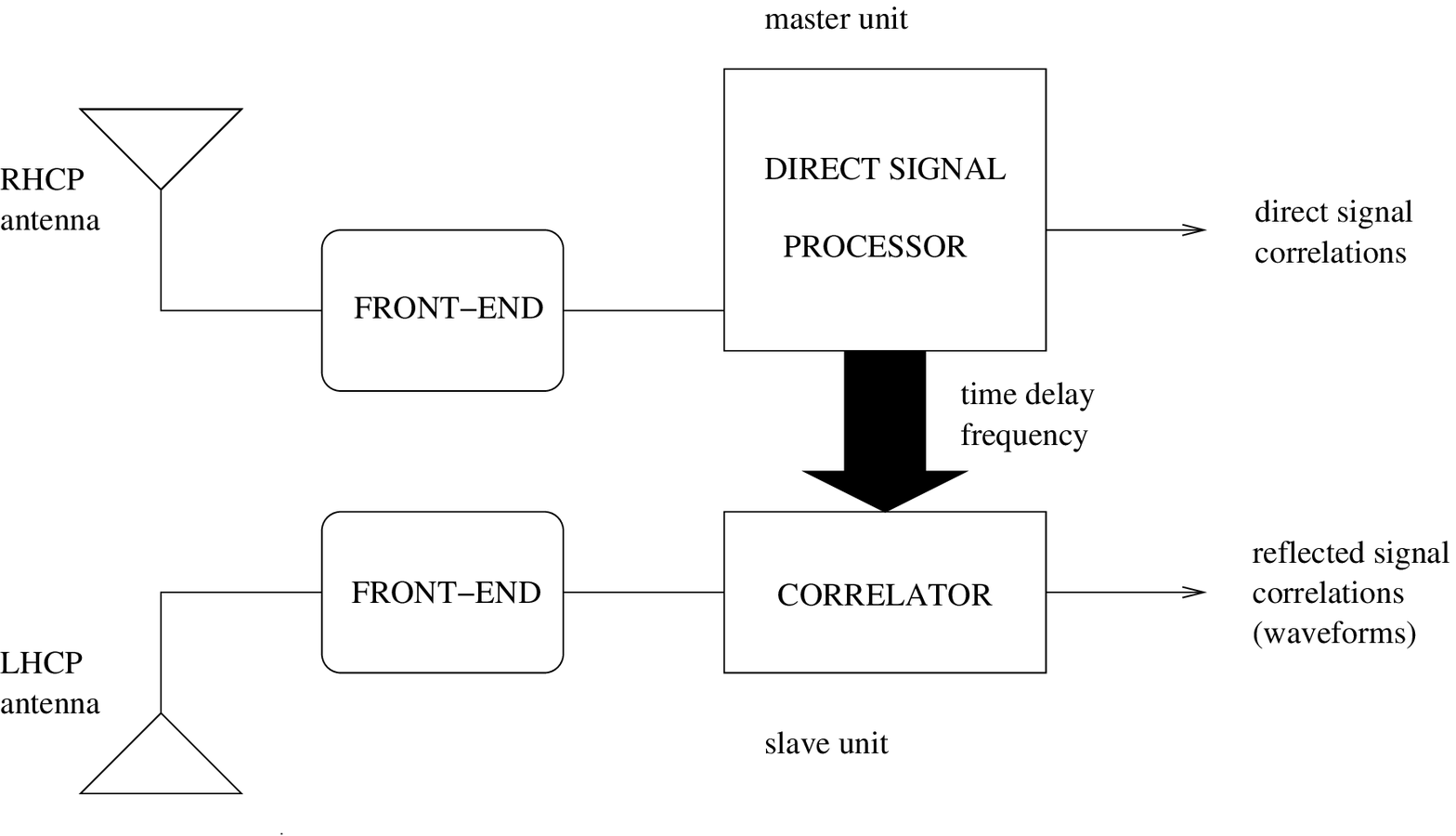} 
    \caption{Simplified block diagram of the GNSS-R tracking concept.}
  \label{fig:track_block}
\end{figure*}
\else
\begin{figure*}[tbhp]
\centering
      \scalebox{.6}{
        \includePSTeX{track_block} 
        }
    \caption{Simplified block diagram of the GNSS-R tracking concept.}
  \label{fig:track_block}
\end{figure*}
\fi

One of the  most relevant tracking parameters is the temporal span of the correlations, i.e.,  the coherent integration time  used to despread the GPS signal. The coherent integration time was set here to 10~milliseconds: it was verified that with this value the ratio of the correlation peak height and the out-of-the-peak fluctuations achieved a maximum. In practical terms, an integration time of 10~milliseconds simplifies the tracking process, as the duration of this time interval is a sub-multiple of the GPS navigation bit duration (with the 50~Hz navigation code in L1). Moreover, we believe that longer correlation times provide some protection from aircraft multipath by mimicking a higher gain antenna (a belief supported by  tests with shorter integration times). 

\section{Retracking the waveforms}
Once a correlation waveform is obtained for both the direct and reflected signals, the lapses  can be estimated. We emphasize that this is not as trivial as considering the maximum sample of each waveform or the waveform centroid, for instance, as the bistatic reflection process deforms severely the signals and, in general, such distortions  will displace  the waveform maximum or centroid location. Moreover, the sampling rate of 20.456~MHz, while  much higher than the Nyquist rate for the C/A code, is equivalent to an inter-sample spacing of 49~ns---about 15~light-meters. This coarseness in the lapse estimation is not sufficient to reach the final altimetric precision target. The main challenge for GNSS-R using the GPS Coarse Acquisition (C/A) code is to provide sub-decimetric altimetry using a 300 m equivalent pulse length, something that can only be achieved by intense averaging with due care of systematic effects. For reference, pulse lengths in monostatic altimeters such as Jason-1 are almost three orders of magnitude shorter.

For these reasons, the temporal position of each waveform (the {\em delay} parameter) is estimated via a Least Mean Squares model fitting procedure. This is  the so-called {\em retracking} process, an otherwise well-known concept in the monostatic altimetric community. The implementation of accurate waveform models (for direct and reflected signals) is fundamental to retracking. Conceptually, a retracking waveform model allows for the transformation of the reflected waveform to an equivalent direct one (or vice-versa), and a coherent and meaningful comparison of direct and reflected waveforms for the lapse estimation.

\subsection{Waveform model}
The natural  model for retracking of the direct signal waveforms is the mean autocorrelation of the GPS C/A code in presence of additive Gaussian noise, which accounts mainly for the receiver noise. As far as the reflected signal is concerned, the model is not so straightforward. In fact, the reflection process induces modifications on the GNSS signals which depend on  sea surface conditions (directional roughness), receiver-emitter-surface kinematics and geometry, and  antenna pattern.  Among all these quantities, the least known ones are those related to the sea surface conditions. In principle, these quantities should be considered as free parameters in the model for the reflected signal waveform and estimated during the retracking process along with the  delay parameter of the waveform.

On the basis of the two most quoted models in the literature for bistatic sea-surface scattering  (\cite{picardi1998} and \cite{zavorotny2000}), we have developed an upgraded waveform model for the reflected signal. This new model, as the two aforementioned, is based on the Geometric Optics approximation of the Kirchhoff theory---that is to say, with a tangent plane, high frequency and large elevation assumption, which is reasonable for the quasi-specular sea-surface scattering at L-band (\cite{soulat2003d}). The main characteristics of this model are the following:
\begin{itemize} 
\item a fully bistatic geometry (as in \cite{zavorotny2000}, but not in \cite{picardi1998}),
\item the description of the sea surface through  the L-band Directional Mean Square Slope\footnote{See \cite{germain2003} for a discussion on the role of wavelength in the definition of DMSS.} (DMSS$_L$) (as in \cite{zavorotny2000}), and
\item the use of a fourth parameter, the significant wave height (SWH), to describe the sea surface (as in \cite{picardi1998}, but not in \cite{zavorotny2000}).
\end{itemize}
We have checked that the impact of SWH mismodeling in our case is negligible,  since the GPS C/A code equivalent pulse width is about 300 meters. Nonetheless, we emphasize that all potential sources of systematic effects must be considered. We foresee a higher and non-negligible impact of SWH if the GPS~P-code (the Precision code) or similar codes in Galileo  are used, since they are an order of magnitude shorter. 

\subsection{Inversion scheme}
The retracking process has been performed through a Least Mean Squares fit of the waveforms with the models described. Because of the speckle noise affecting the reflected signal waveforms, the fit has  not been performed on each complex waveform obtained from the 10~ms correlations. Rather, these waveforms have  first been incoherently averaged: the average of the magnitude of a hundred 10 ms waveforms has then been used to perform the inversion---i.e., 1~second  incoherently averaged real waveforms have been generated for retracking. In this way, reflected/direct temporal lapses have been produced at 1 Hz rate.

In both cases, the fit of the waveform has been performed over three parameters: the time delay,  a scaling factor and the out-of-the-peak correlation mean amplitude.

The geophysical parameters that enter in the model of the reflected signal waveform have not been jointly estimated here. These parameters have been set to some reasonable {\it a~priori} value obtained from other sources of information (Jason-1 for wind speed, ECMWF for wind direction) or from theory (for the sea slope PDF isotropy coefficient). For convenience, we describe the sea surface state using a wind speed parameter, wind direction and the sea slope PDF isotropy coefficient. Using a spectrum model, these parameters can be univocally related to  DMSS$_L$ with the assumption of a mature, wind-driven sea (the sea spectrum in \cite{elfouhaily1997} has been used in this case). We emphasize that DMSS$_L$ is the actual parameter set needed in the reflection model  under the Geometric Optics approximation.

Concerning the reflected signal waveform,  retracking has been performed using only the leading edge and a small part of the trailing edge, since the trailing edge is more sensitive to errors in the input parameters (including geophysical parameters and antenna pattern).


\section{The altimetric algorithm}
The output of the retracking process is the time series of measured lapses. The next step  is finally SSH estimation. In order to solve for this quantity, we have used a differential algorithm: the classical PARIS equation (see~\cite{martin-neira1993}) has not been directly used. Instead, a model for the lapse over a reference surface near the local geoid has been built, and the difference of this reference lapse and the measured one has been modeled as  a function of the height over the reference surface. We call this the {\it differential} PARIS equation (Equation \ref{eq:diffPARIS}). We note that the aircraft INS has been used to take into account the direct-reflected antenna baseline motion and that we have also included both dry and wet tropospheric delays in the model by using exponential models for them with different scale heights and surface values derived from surface  pressure measurements and surface tropospheric delays obtained from ground GPS and SSM/I\footnote{Special Sensor Microwave Imager, a passive microwave radiometer flown aboard United States Defense Meteorological Satellite Program.}). 
\\
%
The {\it differential} PARIS equation writes
\begin{equation}
\label{eq:diffPARIS}
 \Delta_{DM}\;=\;\Delta_D-\Delta_M\;=\;2\;\delta h\;\sin(\epsilon)+b,
\end{equation}
where
\begin{itemize}
\item $\Delta_D$ is the measured lapse, in meters, as estimated from the data,
\item $\Delta_M$ is the modeled lapse, in meters, based on an ellipsoidal model of the Earth, GPS constellation precise ephemeris, aircraft GPS/INS precise kinematic processing, and a tropospheric model, 
\item $\delta h$ is the normal offset between the sea surface  and the (model) ellipsoid surface,
\item $\epsilon$ is the GPS satellite elevation angle at the specular point of reflection, over the ellipsoid, and
\item $b$  is the hardware system bias.
\end{itemize}
The precision obtained after 1-second of incoherent averaging in the estimation of $\Delta_{DM}$  using this approach is displayed in Table 1.  For each PRN number, the root mean squared error of the 1-second lapse is shown (in meters). It is roughly of 3 m. This noise level is as expected from the C/A code bistatic ranging in our experimental setup (antenna gain, altitude, etc.) and consistent with earlier experiments.
\begin{table*}
\vspace{.5cm}
\label{tab:lapses}
\centering{
\begin{tabular}{|r|c|c|}
  \hline
 PRN & Complete track & Beginning of the track \\
        \hline \hline
8  &  3.5 m & 2.7 m\\
  \hline   
24 &  3.4 m & 2.8 m\\
  \hline   
27 &  2.9 m & 2.7 m\\
      \hline
\end{tabular}}
\caption{Precision in the estimation of the time lapses (root mean squared error of the lapses, in meters).}
\end{table*}
\section{Results}
\begin{figure*}[t!]
\centering
        \includegraphics[width=10cm]{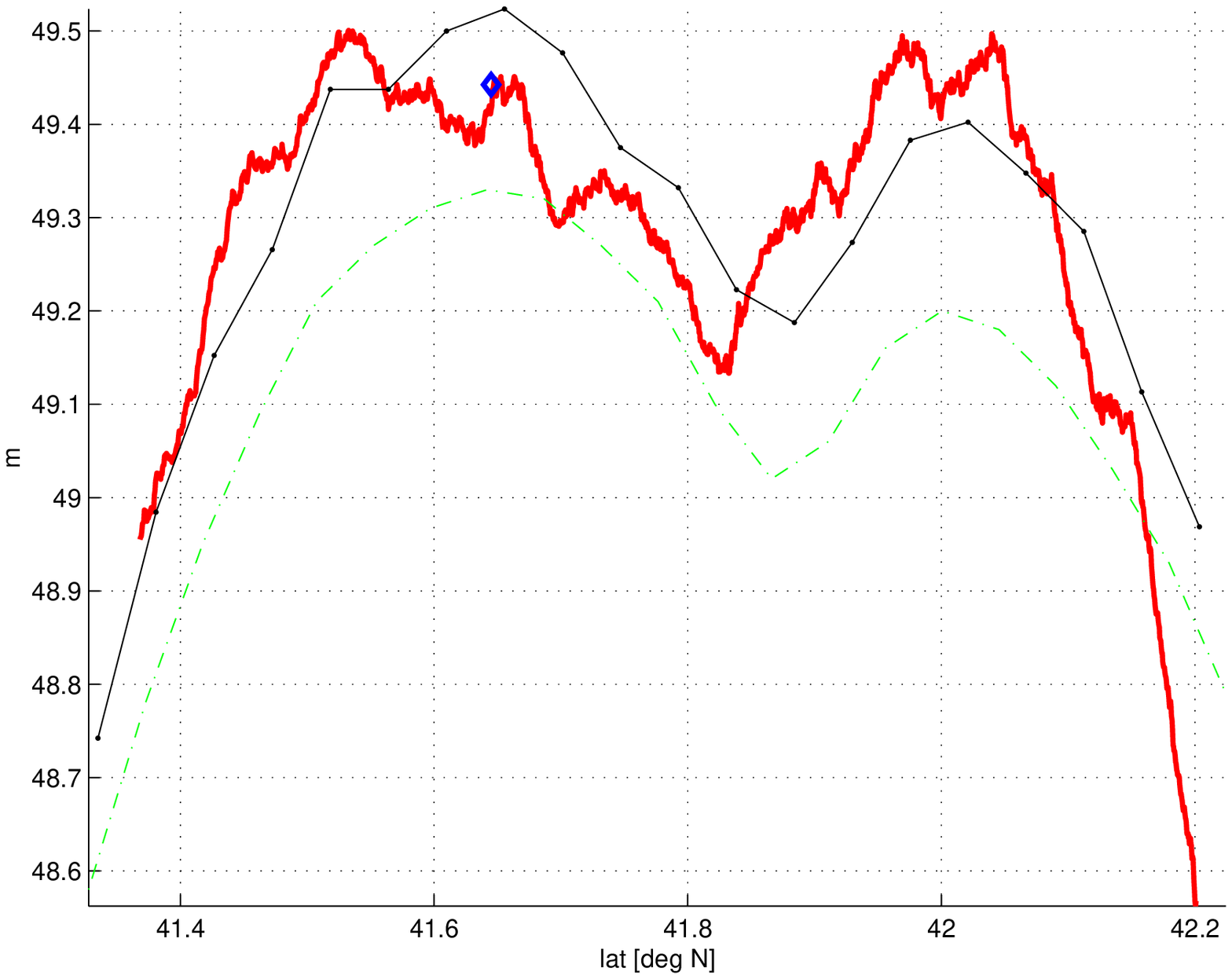} 
    \caption{Final altimetric result. The red line represents the average altimetric result of the three GPS satellite analyzed filtered over a 20~km---i.e., 400~seconds, the aircraft speed being about 50~m/s. The black line represents the Jason-1  SSH, while the green dashed line represents the MSL. The blue diamond is the SSH measurement obtained from the reference buoy.}
  \label{fig:result1}
\end{figure*}

The algorithm outlined  in the previous section has been used with data from the three better behaved GPS satellites. The other two visible satellites appeared to be severely affected by aircraft-induced multipath (probably due to the landing gear and wing surfaces). A SSH profile has been retrieved for each satellite and the average value of the three profiles is shown in Figure~\ref{fig:result1} along with the Jason-1 SSH, Mean Sea Level (MSL) and one SSH value obtained with the control buoy.
This solution has been obtained setting a model wind speed of 10 m/s (values provided by Jason-1 along the track vary between 9 and 13~m/s), wind direction of 0 degrees North (values provided by ECMWF vary between  30 and -20~deg North), and  sea slope PDF isotropy coefficient equal to 0.65 (the theoretical value for a mature, wind-driven sea according to \cite{elfouhaily1997}).
It is important to underline that the use of constant values for the geophysical parameters along the whole track (more than 120~km) induces non-linear errors on the final altimetric estimation. Nonetheless, the bias of the final solution with respect to the SSH (the error mean) is 1.9~cm while the root mean error is 10.5~cm.

\section{Conclusion}

The Eddy Experiment has validated the use of PARIS  as a tool for airborne sea surface height measurements, providing both the  precision and accuracy predicted by earlier experimental and theoretical work. We have observed that the use of a waveform model for the reflected signal, based on geophysical parameters describing the sea surface conditions, is essential for the accuracy of the altimetric solution---a fact which  may  explain earlier results in which no geophysical retracking was performed (e.g., \cite{rius2002}). The accuracy achieved by our algorithm is of the order of 1 decimeter, but we expect that further analysis and refinements, such as the inclusion of DMSS$_L$ parameters in the inversion procedure, will improve these numbers.

Our sensitivity analysis has also shown that the altitude of this flight was not optimal for GNSS-R altimetry and made the experiment more prone to aircraft multipath problems. A higher altitude flight would lead to a smaller angular span of the reflecting area on the sea surface, thus reducing the impact of geophysical parameters, antenna pattern and aircraft multipath on the retracking process of the leading edge,  making the overall altimetric solution more robust. Higher altitudes are also needed to better understand  the space-borne scenario.

We would like to emphasize that GNSS-R signals can be profitably used also for scatterometric  measurements (i.e., {\em speculometry}, from the Latin word for mirror, ``speculo'', see \cite{ruffini2001a}). In a parallel paper (\cite{germain2003}), the inversion of GNSS-R Delay-Doppler Maps for sea-surface DMSS$_L$ estimation is presented for the same data set. The next step is to merge the altimetric and speculometric  processing in an attempt to provide an autonomous GNSS-R complete description of the sea---topography and surface conditions.

We believe the Eddy Experiment is an important milestone on the road to a space mission. We underline that the obtained precision and accuracy are in line with  earlier experiments and theoretical error budgets (see, e.g., \cite{lowe2002b}). We note that the same error budgets  have been used to investigate and confirm the strong impact of space-borne GNSS-R altimetric mission data on mesoscale ocean circulation models (\cite{letraon2003,letraon2003a}). Further analysis of existing datasets (which could be organized in a coordinated database for the benefit of the community) and future experiments at higher altitudes will continue to refine our understanding of the  potential of this technique.

\section*{Acknowledgments}
\addcontentsline{toc}{section*}{Acknowledgments}
This study was carried out under the ESA contract TRP~ETP~137.A.  We thank  EADS-Astrium and all sub-contractors (Grupo de Mecanica del Vuelo, Institut d'Estudis Espacials de Catalunya, Collecte Localisation Satellites, and Institut Fran\c{c}ais de Recherche pour l'Exploitation de la Mer) for their collaboration in the project, and the Institut Cartografic de Catalunya for  flawless flight operations and aircraft GPS/INS kinematic processing. Finally, we thank Irene Rubin from CRESTech, for providing us with SSM/I IWV (Integrated Water Vapor) data. 

{\em All Starlab authors have contributed significantly; the Starlab author list has been ordered randomly.}

\bibliographystyle{plain}
\bibliography{/home/alkaid/intranet/library/feosbiblio.bib}
\addcontentsline{toc}{section*}{Bibliography}
     
\end{multicols}
\end{document}